\documentclass[12pt]{iopart}
\usepackage{graphicx}
\usepackage{dcolumn}
\usepackage{bm}
\def\ddri{\hbox{\rm d}}
\def\Mbf{\mbox{\boldmath $M$}}

\begin{document}
\title[Electronic, optical and thermal properties\ldots]{Electronic, optical and thermal properties of the hexagonal and fcc Ge$_2$Sb$_2$Te$_5$ chalcogenide from first-principle calculations}
\author{T Tsafack$^{1}$, E Piccinini$^{2}$, B-S Lee$^{3}$\footnote{Present address: Tessera Inc., 3025 Orchard Parkway, San Jos\`e, CA, USA}, E Pop$^{4}$  and M Rudan$^{1,2}$}

\address{$^{1}$ DEIS - Dipartimento di Elettronica, Informatica e
  Sistemistica, University of Bologna, Viale Risorgimento 2, I-40136
  Bologna, Italy} 
\address{$^{2}$ E. De Castro Advanced Research
  Center on Electronic Systems ARCES, University of Bologna, Via
  Toffano 2/2, I-40125 Bologna, Italy} 
\address{$^{3}$ Department of
  Materials Science and Engineering and the Coordinated Science
  Laboratory, University of Illinois at Urbana-Champaign, Urbana, IL
  61801, USA} 
\address{$^{4}$ Department of Electrical and Computer Engineering, Micro and Nanotechnology Lab, University
  of Illinois at Urbana-Champaign, Urbana IL 61801, USA }
\ead{enrico.piccinini@unimore.it}
 \date{\today}
\begin{abstract}

We present a comprehensive computational study  on the properties of face-centered cubic and hexagonal chalcogenide Ge$_2$Sb$_2$Te$_5$. We calculate the electronic structure using density functional theory (DFT); the obtained density of states (DOS) compares favorably with experiments, also looking suitable for transport analysis. Optical constants including refraction index and absorption coefficient capture major experimental features, aside from an energy shift owed to an underestimate of the band gap that is typical of DFT calculations. We also compute the phonon DOS for the hexagonal phase, obtaining a speed of sound and thermal conductivity in good agreement with the experimental lattice contribution. The calculated heat capacity reaches~$\sim 1.4 \times 10^6$~J/(m$^3$~K) at high temperature, in agreement with experimental data, and provides insight into the low-temperature range ($<150$ K), where data are unavailable.

\end{abstract}
\submitto{\JPCM}
\maketitle
\section{Introduction}
Over the past two  decades phase-change materials have generated much interest  in the area  of electronic  devices for  memory applications thanks to the scaling  properties, small energy consumption, and large number  of writing  cycles.  The  ability of  such materials  to switch between the  crystalline  and  the amorphous  phase  makes  them  suitable candidates for  data storage.  In  fact the two phases  are associated with  large  differences  in  the optical  constants  and  resistivity~\cite{ovshinsky}. 
Since the  late 1960's digital disk-random access memories (DVD-RAM),  phase-change dual  disks (PD),  re-writable  optical media with increasing storage capability like multilayer DVDs and, later on, solid-state non-volatile memories, have  been designed and released to the market.
\par
Chalcogenide materials like  Ge$_2$Sb$_2$Te$_5$ (GST) have extensively been investigated  either theoretically or experimentally  in order to better understand the nature of their structural, electronic, optical, thermal and electrical properties.
X-ray diffraction  experiments have  provided cell parameters  for the hexagonal  and face-centered-cubic  GST~\cite{petrov,kooi,matsunaga}, and  several  hypotheses  have  also  been made  about  the  amorphous phase~\cite{kolobov, akola, caravati}.
The  GST material  is  a  semiconductor in  both  the crystalline  and the amorphous  phase.   Its optical  band gap  has  been estimated  around 0.5~eV   for  the  former   phase  and   around  0.7~eV   for  the latter~\cite{lee2005}.
\par
In     the     last     decade     several    models     have     been proposed~\cite{pirovano,ielmini,buscemi}  to   describe  the  snap-back phenomenon in  the $I(V)$ characteristic of amorphous  GST glasses. In fact,  such a feature  is fundamental  for using  the material  in the fabrication of solid-state memories.   Even though the hexagonal phase is the  stable one, the metastable  fcc crystals play a  major role in device applications.  As  a matter of fact the  amorphous structure of GST  stems from a  strongly distorted fcc  one~\cite{kolobov}, and the material can easily switch between the amorphous and the fcc phase due  to Joule heating.   The models  describing carrier  conduction in semiconductors are usually based on the knowledge of the electron and phonon dispersion relations for the material at hand.  In a  similar manner this type of data  are useful for a better  understanding  of the  transport  characteristics  of the  GST material.
\par
This paper shows the results of a comprehensive computational study of the  GST  chalcogenide,  including  band  structures  and  optical constants for  both the hexagonal and face-centered  cubic phases. Two former studies devoted to the hexagonal phase were recently published~\cite{lee-jhi,sosso}; they are considered here for comparison purposes. Moreover, the vibrational  properties of  the  hexagonal phase are investigated as well, in order to get information on the speed of sound in the material, on the thermal conductivity, and heat capacity. The  starting point of the  analyses is the  calculation of the band structure by  means of the density-functional  theory using plane waves as  basis set. 
\par
After  calculating the band structure,  the imaginary part  $\epsilon_i  (\omega)$  of  the dielectric  tensor $\epsilon_{\alpha \beta}  (\omega)$ (including  Drude-type contributions)  is derived using  the Drude-Lorentz expression.   The real  part $\epsilon_r (\omega)$    is   then    calculated   through    the   Kramers-Kronig transformation. The Maxwell  model allows  one to link $\epsilon_r  (\omega)$ and $\epsilon_i  (\omega)$ to the  refractive index  $n (\omega)$  and the extinction  coefficient  $k  (\omega)$,  as  well  as  the  absorption coefficient  $\alpha (\omega)$.   Two measurable  quantities  like the optical  reflection  $R(\omega)$   and  transmission  $T(\omega)$  are derived  from $n  (\omega)$  and $k  (\omega)$  using exact  equations considering multiple reflections in a  thin film. They are compared to the corresponding  experimental data.
\par
Finally,  the phonon DOS is calculated  through  the density-functional  perturbation theory (DFPT) for the hexagonal crystalline phase. From this, it is possible to evaluate the sound velocity and the thermal conductivity, which compare well with experimental data on the phonon contribution in hexagonal GST. Moreover, the heat capacity for this phase is obtained over a wide temperature range (5-870 K) by integrating the DOS.

\section{Method and Calculations}\label{method}
The    electronic   structure    has   been    computed    using   the DFT equations  that are implemented in the {\textsc{Quantum  Espresso 4.1}}  code~\cite{qe}.  This software  uses plane waves as  a basis set for the expansion  of atomic orbitals, and implements   periodic   boundary   conditions.   The   local   density approximation (LDA) by  Perdew and Zunger~\cite{perdew} has been considered  for  the  exchange-correlation energy.   The  electron-ion interactions  have been  described by  means of  norm-conserving ionic Bachelet-Hamann-Schluter     pseudopotentials    without    non-linear corrections~\cite{gonze}.   The    valence   configurations   are $4s^24p^2$,   $5s^25p^3$,  and   $5s^25p^4$  for   Ge,  Sb,   and  Te, respectively. Recent papers~\cite{lee-jhi,do} included explicitly the role of Te $4d$ electrons in the valence configuration (and not as a core contribution). Other authors have pointed out that spin-orbit coupling could play a role for such heavy atoms~\cite{lenthe}. As discussed throughout this paper, neglecting these details does not affect the quality of our findings, which favorably compare to experimental evidence.
\par
The cut-off  in the kinetic energy was  set to 80~Ry, a rather conservative choice since preliminary tests proved that changes in  the results  become  less and  less  significative roughly  beyond 50~Ry.
\par
The first step of the  analysis deals with geometry relaxation. As a result of the Born-Oppenheimer approximation, this stage involves the determination of the cell parameters and the atomic  coordinates that minimize the  energy functional  within the  adopted  numerical approximations. 
\par
According to the literature, the stacking sequence of the hexagonal cell is made up of 9 layers. Three possible configurations have been proposed, depending on the position of the Sb and the Ge atoms. In an early work Petrov and coworkers~\cite{petrov} proposed the sequence $\mathrm{Te - Sb - Te - Ge -  Te - Te -  Ge - Te -  Sb}$; more recently, Kooi and de Hosson identified a new stacking where all Sb and Ge atoms exchange their positions~\cite{kooi}, while Matsunaga and coworkers suggested that Sb and Te can randomly occupy the same layer, thus resulting in a mixed configuration~\cite{matsunaga}. Among these configurations, we have adopted that proposed by Kooi and de Hosson, whose total energy is claimed to be the lowest in the computational studies available in the recent literature~\cite{lee-jhi, sosso, sun}. As for the fcc structure, the fact that the phase transition occurs easily between hexagonal and cubic GST suggests that the transformation does not imply a large atomic rearrangement, and the two stackings must share a common background.
\par
\begin{figure}
\begin{center}
\includegraphics[width=0.8\textwidth]{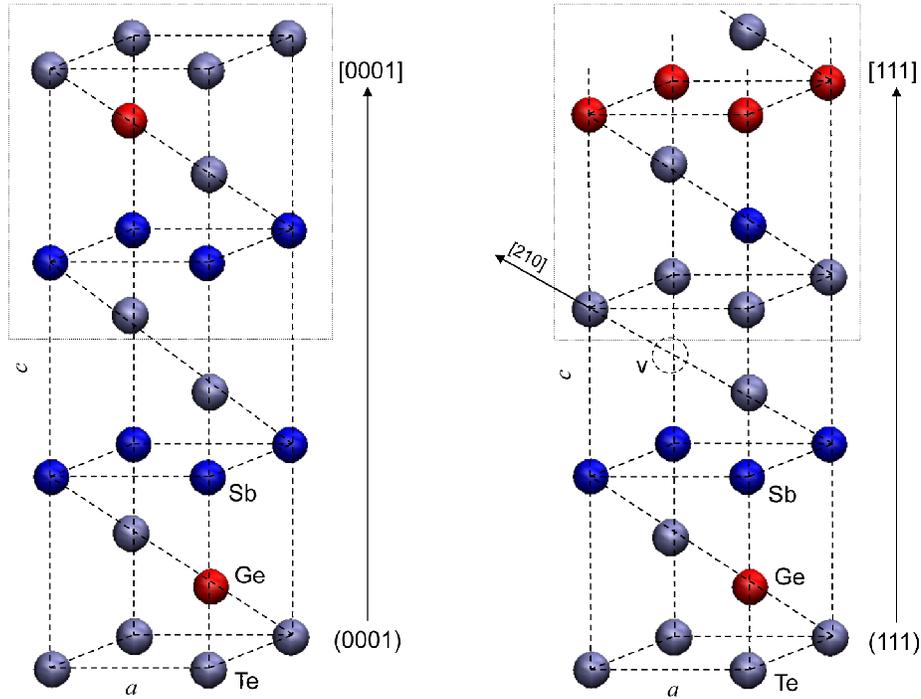}
\caption{Atomic arrangement of the hexagonal (left) 
and fcc (right) GST, showing the stacking sequence 
along the crystalline planes.}
\label{fig_stacking}
\end{center}
\end{figure}

The unit cell for the hexagonal phase here considered is then made up of 9 atoms and arranged in the stacking sequence $\mathrm{Te - Ge - Te - Sb -  Te - Te -  Sb - Te -  Ge}$, while the fcc  structure comes out from  shifting the hexagonal  $\mathrm{Te -  Sb -  Te -  Ge}$ sub-unit along   the  $[210]$   direction to the next crystallographic plane,   thus  creating   a  vacancy   site ($\mathrm{v}$) in between. That leads to a unit cell of $27$ atoms and $3$ vacancies arranged in the  stacking sequence $\mathrm{Te - Ge - Te - Sb  -  Te  -  v  -  Te   -Sb  -  Te  -  Ge}$  repeated  three  times (figure \ref{fig_stacking}).   The  experimental  values  for  the  cell parameters  are:  $a  =  4.22$~\AA,  $c =  17.18$~\AA~for the hexagonal phase~\cite{friedrich},  and  $a_0=6.02$~\AA~\cite{park}, corresponding to $a  = 4{.}26$~\AA,  $c  = 52.13$~\AA~in the equivalent hexagonal system,  for the fcc structure. The geometry relaxation resulted  in a difference from the experimental data of $\Delta  a = 0.08 \%$, $\Delta c =  -3.02\%$ for the hexagonal phase, and  of $\Delta a = -2.05\%$,  $\Delta c = -1.8\%$  for the fcc phase.  Moreover, a slight shift in the position of internal planes is also found.   The calculated shrinkage of the $c$ parameter is consistent with the adopted LDA approximation, and can also be found in the works of Sun \emph{et al.}~\cite{sun} and of Lee and Jhi~\cite{lee-jhi}, but contrasts with the results of Sosso \emph{et al.}~\cite{sosso}. 
\par
A $12  \times 12 \times 4$  $k$-point grid for the  hexagonal GST and, respectively, a  $12 \times 12 \times  1$ grid for the  fcc phase have been used  for the self-consistent  calculation in order  to determine the ground-state configurations for the two systems at hand.
The whole relaxation process for the hexagonal structure took around 2 days on a 8-processor Linux  cluster.  Due to the intrinsically higher structural complexity, the computational  load for the fcc cell proved to be $4$ times higher.
\par
As the  material optical response  is due to transitions  within and between  valence and  conduction  bands, the  first  step towards  its calculation,  once the  ground state  is known,  involves  computing the eigenfunctions and eigenvalues also  for the conduction band. A uniform grid of  $20\times 20\times 20$  was used at  this stage for  both the hexagonal and the fcc cases.  As  the optical response strongly depends on the transitions  to the conduction band, introducing a  dense grid  in the calculations  increases the accuracy  of the  calculations themselves. The equations used to  build the complex dielectric tensor $\epsilon_{\alpha \beta} (\omega)$ are reported in the appendix.
\par
The last  part of the  present investigation concerns  the vibrational modes.   To  this  aim  we  have  adopted  the DFPT approach~\cite{baroni} provided by the {\textsc{Quantum Espresso}}  package. This method sidesteps the need  of   constructing  a   superlattice  typical  of   the  standard frozen-phonon   framework~\cite{parlinski},  and  allows   one  to calculate the phonon-dispersion relation.  The calculation breaks into three  steps,  namely, \emph{(i)}  computing  the ground-state  charge density  for  the   unperturbed  system,  \emph{(ii)}  evaluating  the phonon frequencies and the  dynamical matrices at a given $q$-vector and, \emph{(iii)} transforming the dynamical matrices back in the real space.   The  calculation  of   the  ground-state  charge  density  is performed  by the  self-consistent procedure  described  earlier.  The parameters  used  in   step  \emph{(i)}  (cutoff  energy,  convergence threshold, Gaussian smearing, and so on)  are the same as those of the band-structure  calculation.  However,  a $4\times  4  \times 1$-dense $k$-point  grid has  been  adopted here.   The  phonon calculation  is performed with a $4\times 4 \times 4$ $q$-vector grid.
\section{Results and Discussion}
\subsection{Band Diagram and Density of States}
\begin{figure}
\begin{center}
\includegraphics*[width=\textwidth]{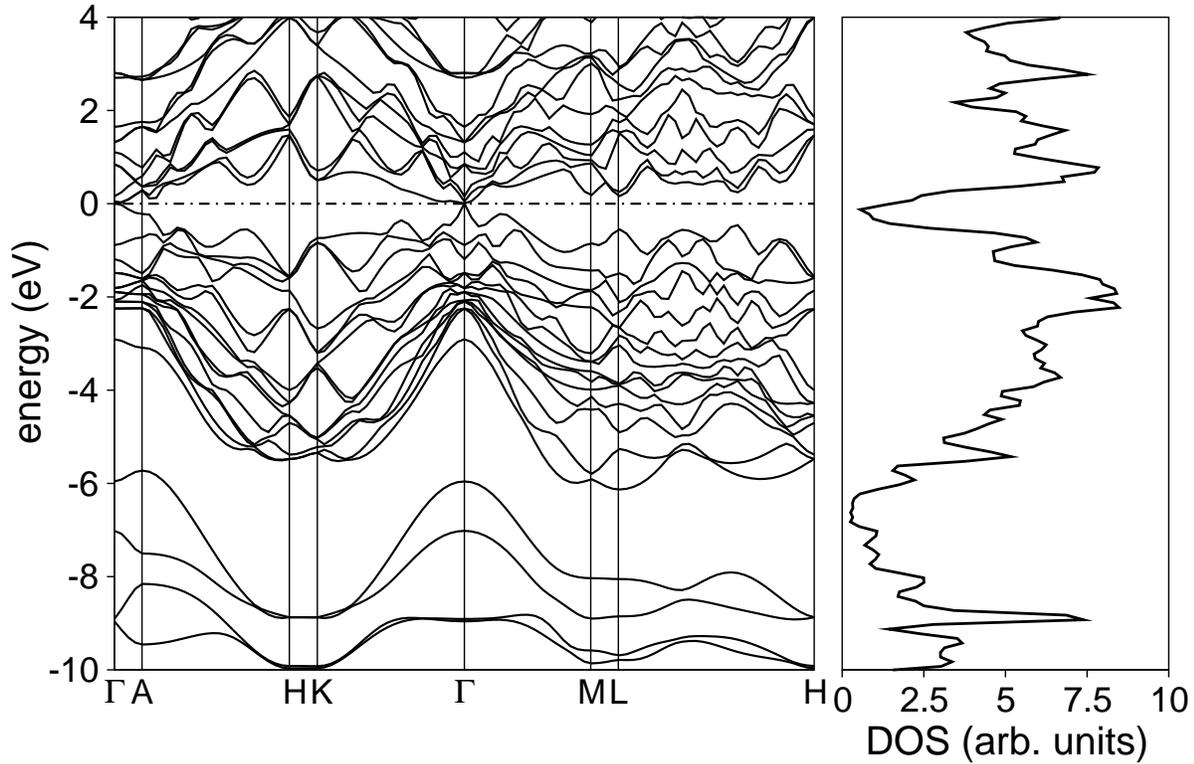}
\caption{Band diagram for the hexagonal phase along the high symmetry
lines (left), and corresponding DOS (right).
The predicted Fermi level is located at 0~eV.} 
\label{hex_bdndos}
\end{center}
\end{figure}

\begin{figure}
\begin{center}
\includegraphics*[width=\textwidth]{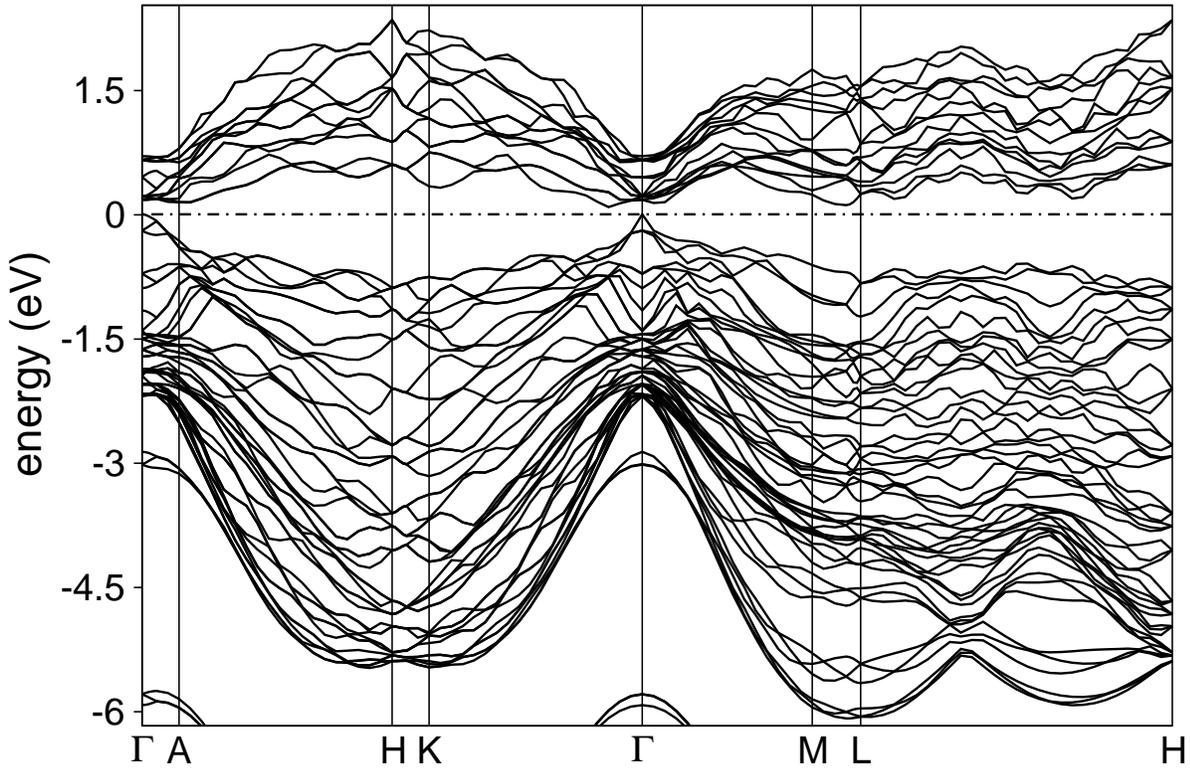}
\caption{Band  diagram  for the  fcc  phase  along  the high  symmetry
lines. The predicted Fermi level is located at 0~eV. Only the valence band and the bottom of the conduction band are shown. An indirect bandgap of $\sim0.1$~eV is found along the $\Gamma$--K line; the energy gap at $\Gamma$ is about 0.2 eV. The corresponding DOS is shown in figure {\protect{\ref{fcc_dos}}}.}
\label{fcc_bnd}
\end{center}
\end{figure}

In figures~\ref{hex_bdndos}  and \ref{fcc_bnd} we  report the electronic band structures  along high-symmetry lines  around the  top of  the valence band (VB) and  the bottom of the conduction band  (CB). The  DOS  is also  shown.  The  actual calculation  was performed  in an energy interval larger than that  shown, this proving the existence of a few deeper bands. Apart from the extension of the band gap that will be discussed later, the shape of the bands compares favorably with the calculations of Yamanaka \emph{et al.}~\cite{yamanaka} and, despite the different parametrization of the pseudopotentials, matches very well the results by Lee and Jhi~\cite{lee-jhi}, both qualitatively and quantitatively. A preliminary band diagram for the fcc  phase has recently been   published   by  some   of   the   authors ~\cite{piccinini}.
\begin{figure}[h]
\begin{center}
\leavevmode
\includegraphics*[width=\textwidth]{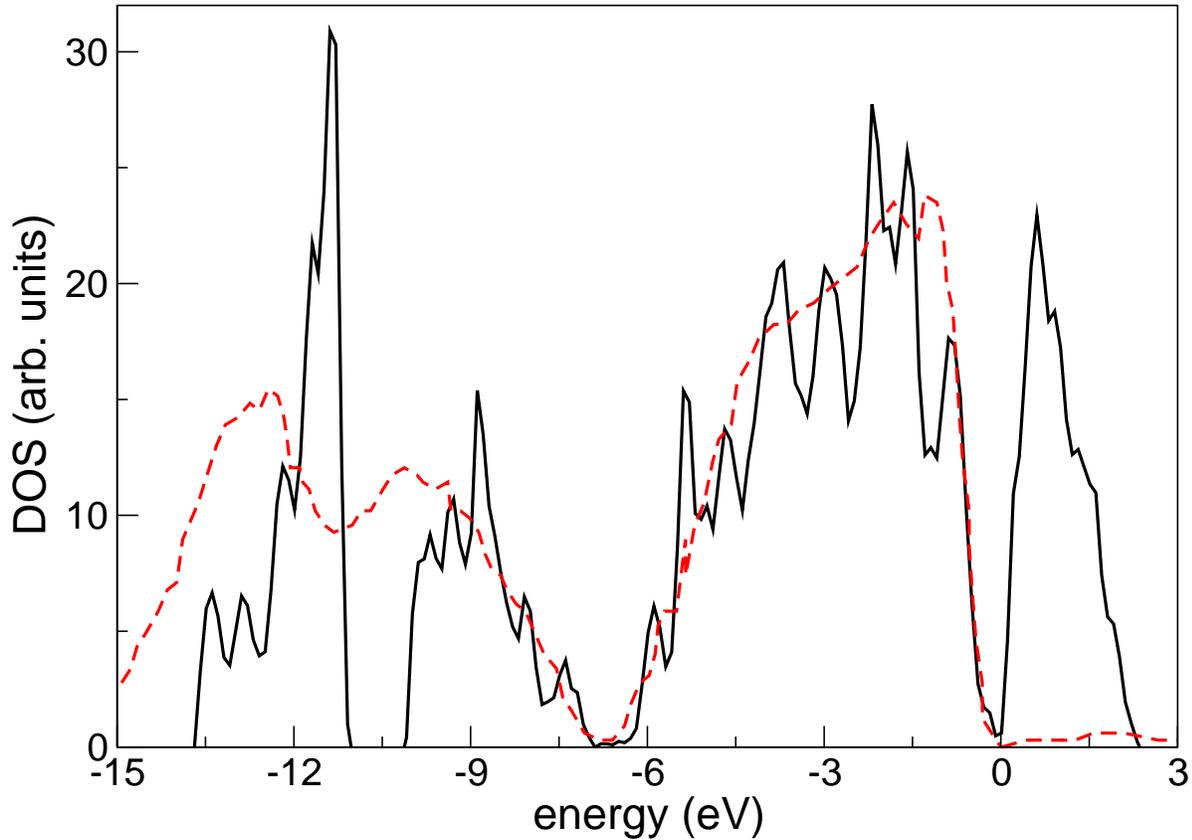}
\caption{Comparison between calculated       (black,      continuous       line)      and
experimental~{\protect{\cite{kim}}} (red,  dashed line) densities of
states for the fcc phase. The  non-negligible value for  the DOS in  the band gap (around $E$=0~eV) present in the calculated curve is an artefact due to the smearing of the interpolating Gaussian function.}
\label{fcc_dos}
\end{center}
\end{figure}
\par
As a result of the simulations,  a band gap smaller than what measured in  optical  experiments  (0.5~eV)~\cite{lee2005, shportko} is  found  in  both  cases.   More specifically, the hexagonal phase  apparently acts as a semi-metal (VB and CB  are degenerate at  the $\Gamma$ point),  whereas an indirect band gap of about 0.1~eV is found for the fcc phase. This result is consistent in shape with the findings of optical experiments that indicate an indirect bang gap for this phase.
\par
In the  recent  works of Lee and Jhi~\cite{lee-jhi} and of Sosso \emph{et al.}~\cite{sosso} a band gap of about 0.2~eV, smaller than the optically-determined one, is found also for the hexagonal phase. The work of Lee and Jhi and that of Sosso \emph{et al.}\ do not share the same parametrization of the valence electrons for Te, nor have the same size of the unit cell, but achieve similar results for the band gap. On the other hand, the shape of the bands found in this work is almost the same as that of Lee and Jhi and, once the conduction band obtained by our calculation is shifted towards higher energies, it can be superimposed almost exactly to that of Lee and Jhi. Moreover, apart from high-frequency oscillations probably related to different interpolating schemes, the calculated DOS for the hexagonal phase is consistent with that of Sosso \emph{et al.}\ for both the valence and conduction bands. The same situation also holds true for the fcc phase with respect to experimental data (figure~\ref{fcc_dos}). One difference between this result and those of Sosso \emph{et al.}\ and of Lee and Jhi relies on the approximation of the exchange-correlation potential (LDA instead of the generalized-gradient approximation). The use of different parametrizations for the pseudopotentials and the exhange-correlation term results in different lattice constants and band gap values. Nevertheless, the discrepancies in the band gap among this work and the two references above are well within the intrinsic procedure error~\cite{gygi}.
\par
The underestimation of the band gap is a well known effect of the DFT calculation and can be corrected by the GW approach and the Bethe-Salpeter equation, to take into account many-body effects~\cite{onida}. 
\par
Despite this limitation, DFT is able to reproduce trends, such as a variation in the band gap due to structural changes. This is the case of the slight increase in the band gap found in the transition from the hexagonal to the fcc phase. In fact, the stoichiometry of the fcc phase implies that 20\% of the lattice positions are represented by vacancies, situated between two well-defined sub-units of the unit cell. Due to the increased distance, the Te-Te bond of the fcc structure is much weaker than that of the hexagonal counterpart. When a melt is quickly undercooled to the amorphous state, the number of weak bonds found in the final structure is quite large, and rings and structural defects are also found~\cite{akola,caravati,hegedus}. According to the capability of predicting trends of the DFT calculations, since the entropy grows from the hexagonal to the fcc crystal and from the fcc phase to the amorphous one, a wider band gap is expected for the latter, consistently with optical determinations.  For these reasons, the obtained bands are suitable for being incorporated into a transport simulation scheme that takes into account all of the material phases, including the amorphous one.
\par
The  second  effect  leading  to  the underestimation of the band gap  is that the measured band gap depends on  the  position of  the  Fermi  level.   For a  $p$-type  degenerate semiconductor  such as  crystalline  GST~\cite{leebook}, the  Fermi level is inside  the valence band. As a  consequence, for an interband optical  transition to  occur, a  photon  must be  absorbed having  an energy larger than the  difference between the band edges.  Therefore, the optical band gap of  a degenerate semiconductor is larger than the electronic band  gap (Burstein-Moss shift~\cite{mossbook}).   A proof that the crystalline GST is  a $p$-type degenerate semiconductor comes from the  experiments based  on the  Hall  effect. Indeed,  to explain  the temperature-dependence  of the  Hall  coefficient it  is necessary  to assume  that the Fermi  level for  the hexagonal  GST is  about 0.1~eV lower than the valence band edge~\cite{lee2005}.
\subsection{Optical Properties}
\begin{figure}[h]
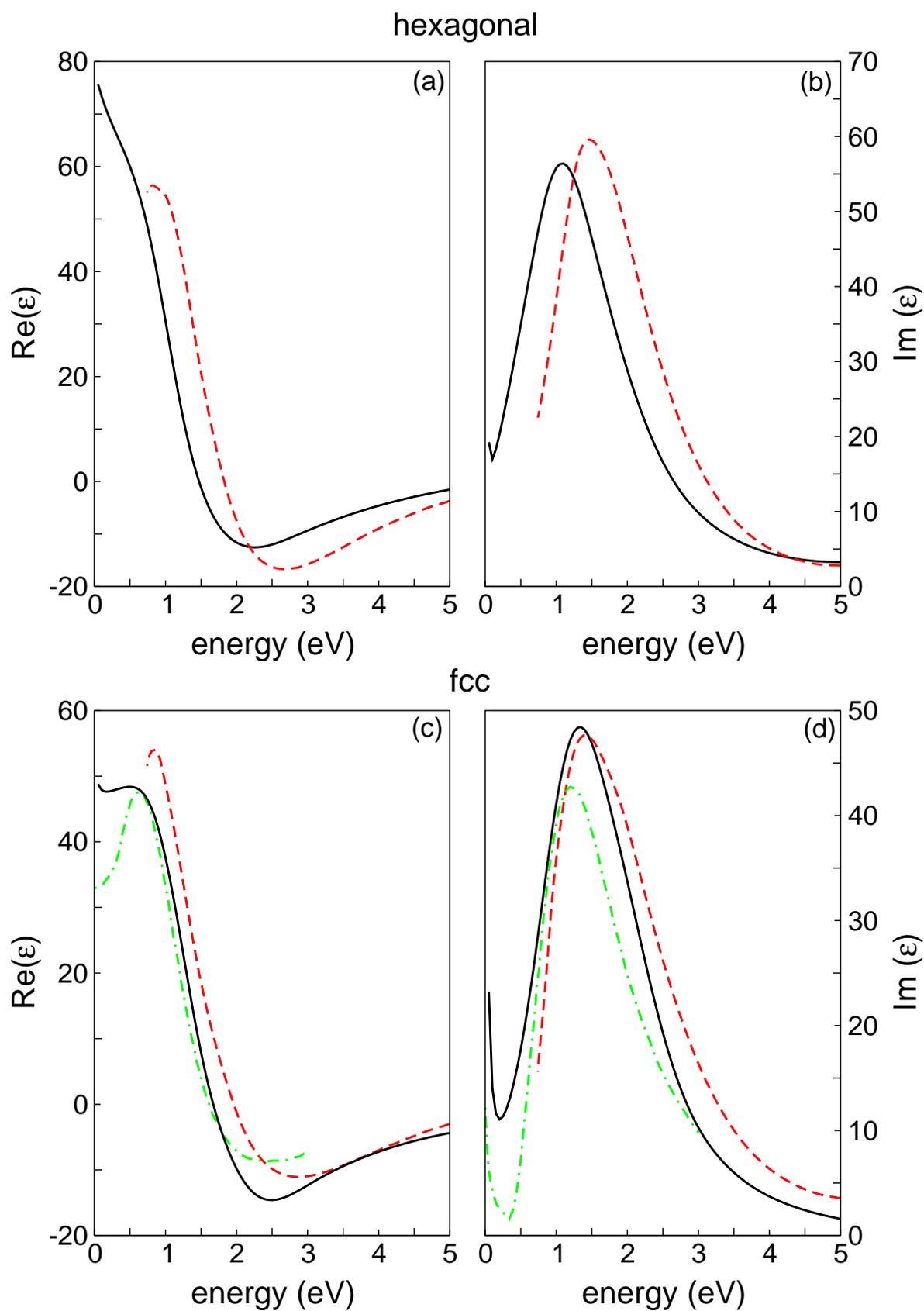

\begin{center}
\includegraphics*[width=\textwidth]{figure5a_5b.eps}
\includegraphics*[width=\textwidth]{figure5c_5d.eps}
\caption{Real part ((a) and (c)) and Imaginary part ((b) and (d)) of the dielectric  function for  the  hexagonal and  the fcc phases. The (red) dashed~\cite{lee2005,leebook} and (green) dash-dotted lines~\cite{shportko} show the corresponding functions derived from optical measurements.}
\label{epsilon}
\end{center}
\end{figure}

\begin{figure}[h]
\begin{center}
\leavevmode
\includegraphics*[width=\textwidth]{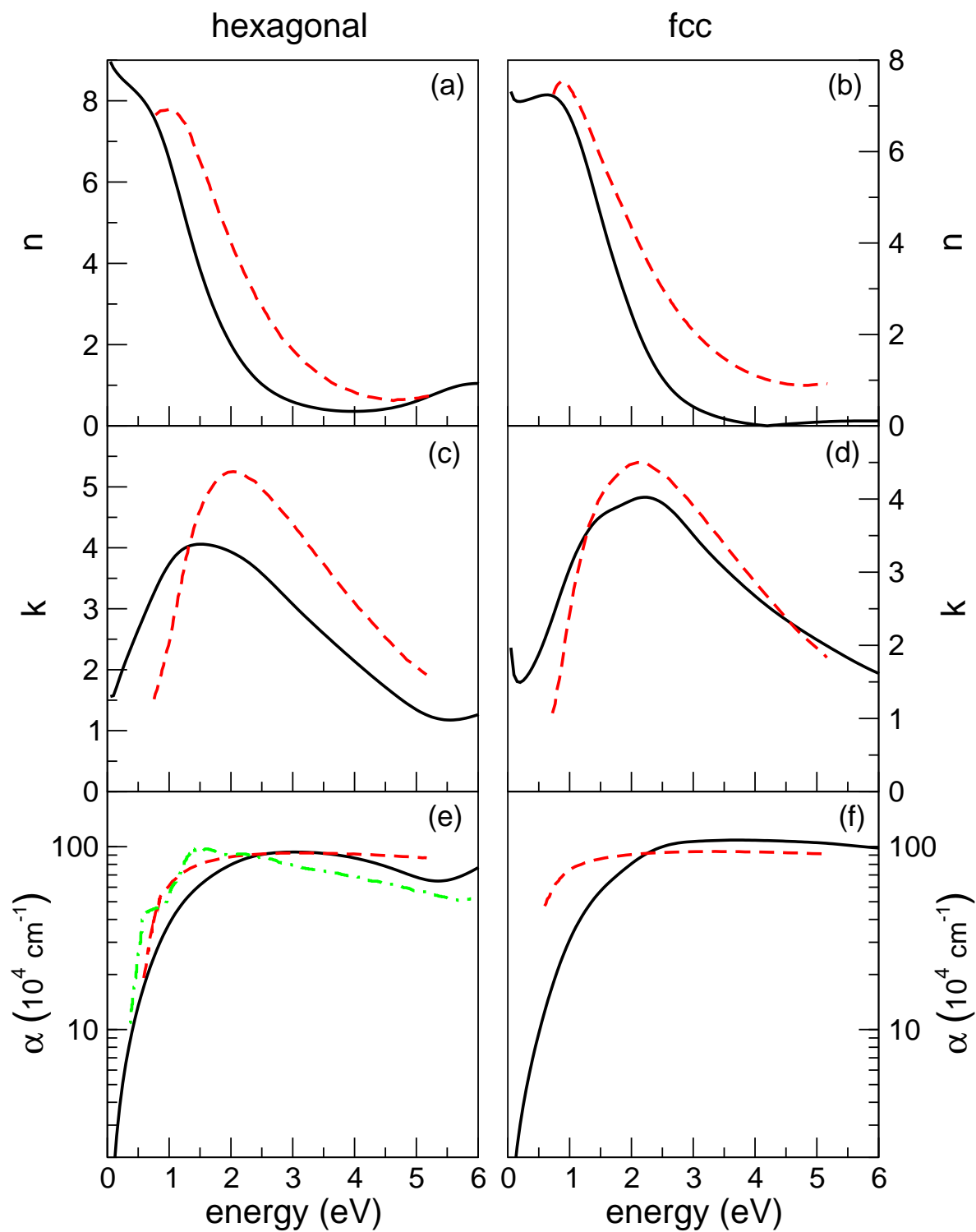}
\caption{Refractive index ((a) and (b)), extinction coefficient ((c) and (d)), and absorption coefficients ((e) and (f)) for the hexagonal and fcc phases. The (red) dashed line shows experimental data from~\cite{lee2005}; the (green) dash-dotted line in (e) is taken from~\cite{lee-jhi}. Please note that in the original paper the latter curve has been shifted towards higher energy to correctly reproduce the band gap.}
\label{optical}
\end{center}
\end{figure}

The  calculated  real and  imaginary  parts, $\epsilon_r(\omega)$  and $\epsilon_i(\omega)$, of the dielectric function of the two phases are shown   in  figure~\ref{epsilon}.   They   are  superimposed   with  the corresponding     experimental      relations     found     in     the literature~\cite{lee2005,shportko,leebook}.  To  properly  compare  the experimental and theoretical  data it is necessary to  remind that the dielectric function depends on  the band gap. The detailed expressions are shown in the appendix.   As the DFT calculation underestimates the band gap, we expect that the calculated dielectric function be rigidly shifted on the energy axis toward the lower energies with respect to the experimental one.  This indeed happens, and the horizontal  offset found between the  experimental and theoretical curves (approx.  0.5~eV) complies with such an interpretation. Since DFT does not take into account many-body effects, excitonic effects have been ignored.
The calculated  and experimental refractive index $n(\omega)$,  along with the extinction and absorption coefficient  $k(\omega)$, $\alpha(\omega)$ are also compared (figure~\ref{optical}). Similar calculations for the absorption coefficient are available in the literature~\cite{lee-jhi} for the hexagonal phase, though using a different set of pseudopotentials, and are reported in figure~\ref{optical}(e) for a straightforward comparison.
As    $n(\omega)$,    $k(\omega)$   and $\alpha(\omega)$  are   calculated  through  $\epsilon_i(\omega)$  and $\epsilon_r(\omega)$,   the   same    reasons   accounting   for   the discrepancies in the dielectric function still hold true.  However, we stress a better matching for the fcc data, which may be an evidence of a calculated band gap closer to the experimental one.
\par
It is also worth noting that the optical determination of the band gap requires  extra calculations.   In fact,  as shown  in the  two bottom panels of figure~\ref{optical},  the absorption coefficient $\alpha$ can be  measured accurately  only in a range of  energies that  is somewhat larger than the optical band  gap.  As a consequence, the intercept of the experimental $\alpha (E)$ curve with the energy axis must be found by  extrapolation.  This is  typically done  by assuming  a power-like relation~\cite{mottbook}
\begin{equation}\label{parabolicmodel}
\alpha h\nu \propto {(h\nu - E_g^{\rm opt})}^r \,,
\end{equation}
where $h\nu$  denotes the photon  energy, $E_g^{\rm opt}$  the optical band gap,  and the exponent $r$  equals $2$ for an  indirect band gap. The  value of  $E_g^{\rm opt}$  is determined  by the  intersection of $(\alpha  h\nu)^{1/r}$ with  the  energy axis  $h\nu$.  However, equation (\ref{parabolicmodel})  relies  on   a  model  which  simplifies  the calculated  bands.  This  introduces  another  error  source  in  the determination  of  the band  gap,  that  adds  to the  ones  discussed earlier.
\par
To better compare numerical results with experiments it is also useful to  calculate  practically   measurable  quantities  such  as  the optical transmission $T(\omega)$ and reflection $R(\omega)$.  As in most cases GST samples are available as thin films on substrates, it is necessary to account  for the dependence  of $T(\omega)$ and $R(\omega)$  on the film  thickness.    Two  GST  samples   with  significantly  different thicknesses have been  prepared and tested for each  phase.  They were sputter-deposited  on  glass slides  and  then  annealed  in an  argon atmosphere for $20$ minutes at  $180^{\circ}$ C (for the fcc phase) or $360^{\circ}$ C  (for the  hexagonal phase).  Following  the procedure described   elsewhere,~\cite{shportko}   the   optical   transmission $T(\omega)$ and  reflection $R(\omega)$ were measured  at an incidence angle of $0^{\circ}$ and $7^{\circ}$, respectively.  The  optical   thickness,  estimated  by  fitting  the   data  to  the previously-obtained optical  constants, is 15  and 240~nm for  the fcc samples, and 12 and 240~nm for the hexagonal samples.
\par
The  dependence  of $T(\omega)$,  $R(\omega)$  on  thickness has  been evaluated      numerically       by      solving      the      optical equations~\cite{tsu} for  a normally-incident light  on a thin layer on top  of a thick glass substrate with $n  = 1.5$ and $k = 0$.
\begin{figure}[h]
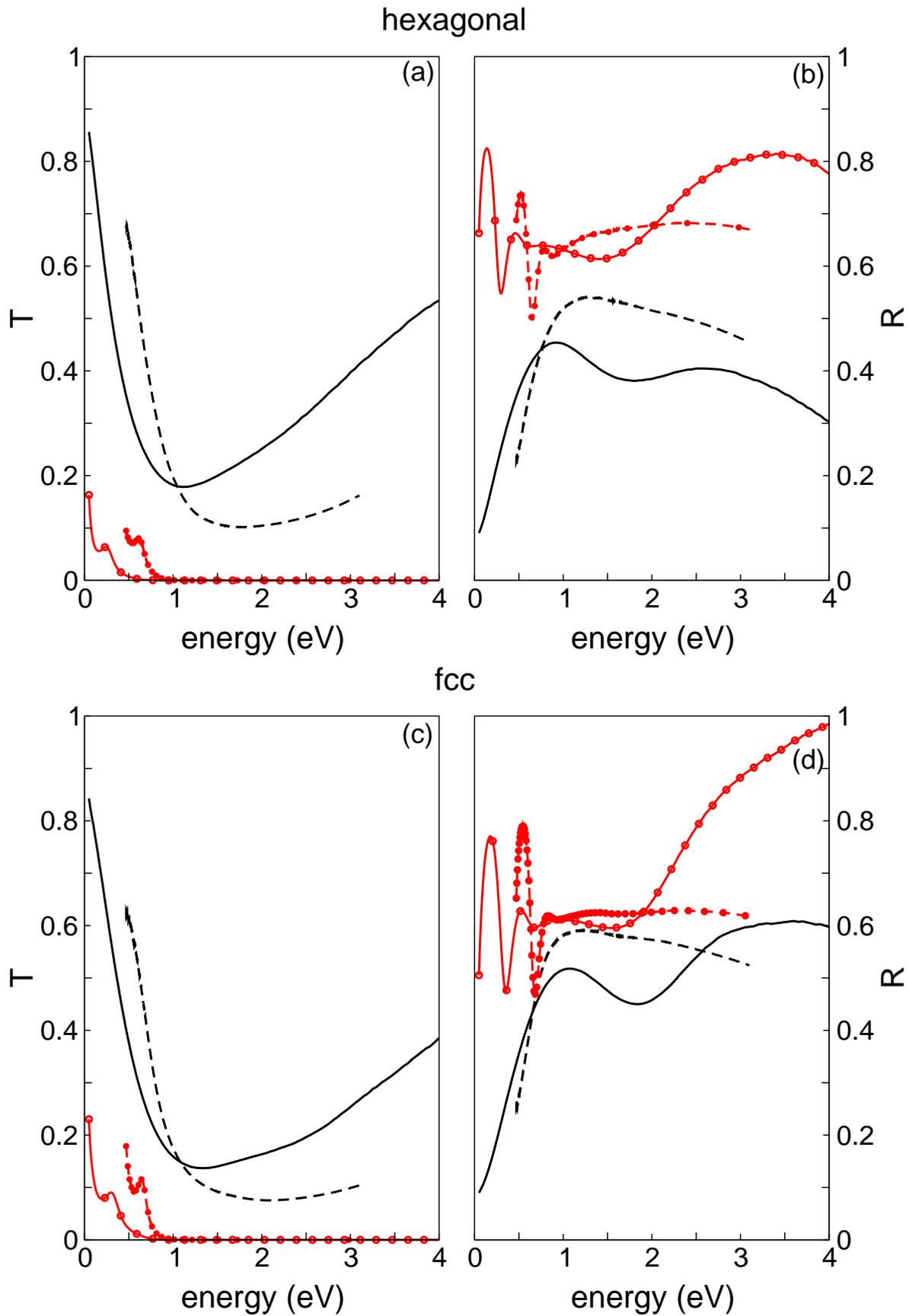

\begin{center}
\includegraphics*[width=\textwidth]{figure7a_7b.eps}
\vskip 1mm
\includegraphics*[width=\textwidth]{figure7c_7d.eps}
\caption{Optical transmission  $T$ ((a) and (c)), and reflection $R$ ((b) and (d)) for a thin (12 or 15~nm, black lines) and for a thick sample (240~nm, red lines with solid dots). Calculated values are represented by continuous lines, while dashed lines refers to experiments.}
\label{hexfcctrcalvsexp}
\end{center}
\end{figure}
The results are reported in figure~\ref{hexfcctrcalvsexp}. In both cases the transmission $T(\omega)$  scales down and, conversely, $R(\omega)$ scales  up with  thickness, as  should be.   Interference  fringes are present  in the  spectra near  or below  the optical  band  gap, since multiple  reflections occur inside  the film  and interfere  with each other.   Once  again,  the  calculated  data suffer  from  the underestimation  of  the  band   gap,  but  the  comparison  is  quite satisfactory, especially for the thick samples.
\subsection{Phonon calculation}
\begin{figure}[h]
\begin{center}
\includegraphics*[width=\textwidth]{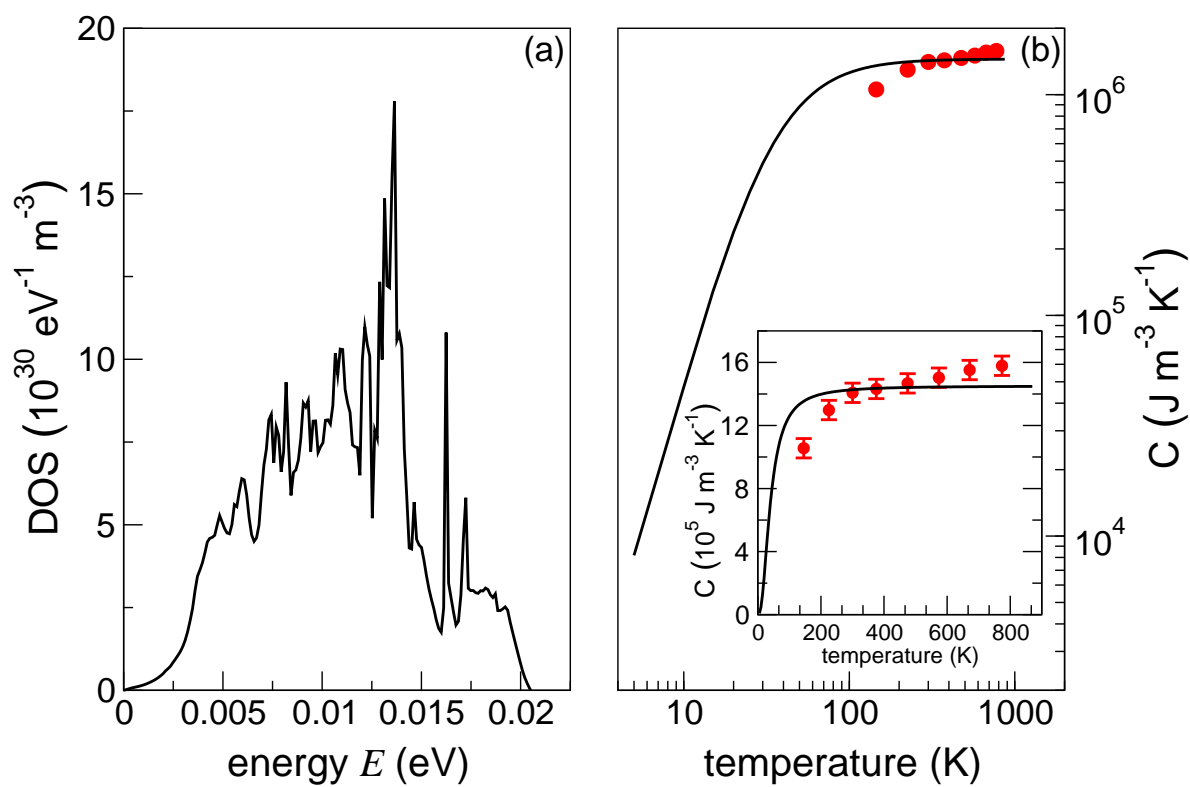}
\caption{(a) Phonon density  of states for the hexagonal phase. (b) Heat capacity for hexagonal GST calculated using data from the panel (a) and equation~(\ref{HeatCapacity}) (black line) compared to experimental data (solid red dots) taken from ref.~\cite{kuwahara}. The inset shows the same data in the linear scale to better represent the region above room temperature, and the error bars for the experimental data.}
\label{hexphndoshc}
\end{center}
\end{figure}

The calculation of the full dispersion spectrum is a rather demanding task, and very strict convergence criteria are often required. Therefore, we have limited our analysis to the DOS with the aim of calculating the speed of sound and heat capacity of the material, which can be directly compared with experimental data. More details about the complete phonon spectrum are left to future work.
The phonon DOS for the hexagonal GST is shown in figure~\ref{hexphndoshc}(a). The general tendency of chalcogenides to have very low phonon frequencies in the range of few tens of meV~\cite{sosso} is confirmed by our findings. The analogous calculation for the fcc phase resolved into unstable results and a number of imaginary frequencies were also found with any reasonable set of the simulation parameters cited in section \ref{method}. This calculation is omitted from  the present publication; however this  may prove once more that the fcc structure is metastable.
\par
The obtained speed of sound along the three orthogonal directions is around $v_{t1}=1.74$~nm/ps, $v_{t2}=2.24$~nm/ps  and $v_l=3.36$~nm/ps for the two  transverse and the longitudinal branch, respectively. The last value compares well with the experimentally estimated $\sim 3.3$~nm/ps       reported      in       the literature~\cite{lyeo}. In the  high temperature limit, the speed  of sound  can be  exploited  to determine  the lattice contribution to the minimum  thermal conductivity $\Lambda_{\rm min}$ of the material:
\begin{equation}\label{MinimumConductivity}
\Lambda_{\rm min} = \frac{1}{2}
\left(\frac{\pi}{6}\right)^{1/3}k_Bn^{2/3}(v_l+v_{t1}+v_{t2}) \,,
\end{equation}
where $n \approx 3.4\cdot10^{22}$~cm$^{-3}$  is the atomic density, and $k_B$ is the Boltzmann constant. The lattice contribution to the minimum thermal conductivity  is  $\Lambda_{\rm min}=0.43$~W/(m K), a lower value than  those  observed  in experiments for  the hexagonal phase. 
\par
However, this result must be interpreted with care, and three aspects deserve attention.  First, it should be pointed out that the hexagonal phase is the only stable phase existing at high temperatures (typically above 600~K), and data often refer to that range.
\par
Next, according  to   Reifenberg  and co-workers~\cite{reifenberg}, the  GST thermal conductivity depends also on the film thickness.  For  the hexagonal phase they found a decrease from $1{.}76$~W/(m K)  for a $350$-nm thick sample  to $0{.}83$~W/(m K) for  a   $60$-nm  thick  sample.    %
\par
Finally, the carrier density in hexagonal crystalline GST is relatively large and electrical carriers also contribute to the heat transport. Experiments have estimated that the electrical contribution is roughly equivalent to the lattice contribution~\cite{yang}, thus leading to an overall conductivity about twice that of $\Lambda_{\rm min}$ calculated above. Thus, taking into account these remarks, $\Lambda_{\rm min}$ is consistent with the phonon contribution in the experiments.
\par
A further confirmation about the validity of the reported DOS comes from a comparison of the calculated heat capacity of GST with that experimentally-determined by Kuwahara and co-workers~\cite{kuwahara}. Let $E=\hbar \omega$ be the energy of the phonon; the heat capacity can be calculated from the simulated phonon DOS by means of:
\begin{equation}\label{HeatCapacity}
C=\int_0^{\infty} E \, \frac{\partial f_{BE}(E)}{\partial T} \, \mbox{DOS}(E) \ddri E \,
\end{equation}
where $f_{BE}(E)={\big\{\exp[E/(k_BT)]-1\}}^{-1}$ is the Bose-Einstein distribution function, and $T$ is the temperature. The calculated and experimental data are reported in figure~\ref{hexphndoshc} up to 870 K, which corresponds to the approximate melting temperature of GST. According to Kuwahara, the experimental heat capacity slightly increases in the high-temperature region, as the result of structural relaxation of point defects. However, the integral in~(\ref{HeatCapacity}) includes only the lattice contribution to heat capacity, and thus predicts a saturating value in the classical limit at high temperature. Nevertheless, the comparison is good, and calculated data are consistent with experiments in the whole range examined. In addition, these calculations provide an estimate of the heat capacity of hexagonal GST in the temperature range where experimental data are unavailable ($T<150$~K).
\section{Conclusion}
In  this paper, we  reported the  electronic and optical properties  for the hexagonal  and face-centered  cubic phases  of the Ge$_2$Sb$_2$Te$_5$ chalcogenide.
\par
The electronic band diagram and DOS were calculated using the density functional theory combined with planes waves, norm-conserving pseudo-potentials and the local density approximation implemented in the code \textsc{Quantum Espresso}. The band diagram and DOS for the hexagonal phase are in good agreement with those reported in the literature. Even though DFT equations are known to underestimate the band gap, the shape of the bands confirms the existence of an indirect band gap for the fcc phase, and the DOS of the latter correctly compares to previously published data. The calculation also showed a tendency of the band gap to increase with respect of the degree of disorder of the cell. This result makes the band diagrams suitable to be used in transport simulations that describe the electrical behaviour of GST.
\par
The dielectric function was obtained implementing the Drude-Lorentz expression and the Kramers-Kronig relationships. Furthermore, the refractive index, the extinction and absorption coefficients were derived from the Maxwell model. By incorporating these functions into equations including multiple internal reflection, the optical transmission and reflection for a thin chalcogenide film deposited on a glass substrate were calculated and then compared to experiments. Most of the differences in the comparison can ascribed to the underestimation of the band gap. 
\par
Moreover, the density functional perturbation theory allowed us to calculate also the phonon DOS for the hexagonal phase. The analysis of the acoustic modes for the hexagonal phase led to reasonable values for both the speed of sound and the minimum thermal conductivity at room temperature. The heat capacity from 5 K up to the melting temperature is also presented, in good agreement with experimental data at high temperature, and providing insight into the low temperature range ($T<150$~K) where data are unavailable.

\ack
The authors appreciate the helpful ideas and suggestions from Prof.\ C.\ Jacoboni and Prof. R.\ Brunetti (University of Modena and Reggio Emilia), Prof.\ P.\ Giannozzi (University of Udine), and Dr.\ S.\ Pamidighantam (National Center of Supercomputing Applications at the University of Illinois). The authors also want to credit Dr.\ A.\ Calzolari and Dr.\ A.\ Ferretti (University of Modena and Reggio Emilia) for their support in the start-up of the simulation environment. A special thank is also due to Prof. J.R. Abelson and Prof. S.G. Bishop for useful discussions and support.
\par
Bong-Sub Lee was supported by the National Science Foundation under Awards No. DMR 07-06267.  The experimental samples were supplied by Dr.\ Byung-ki Cheong at Korea Institute of Science and Technology, and the optical properties were measured in the Frederick Seitz Materials Research Laboratory Central Facilities, University of Illinois, which are partially supported by the U.S.\ Department of Energy under Grants DE-FG02-07ER46453 and DE-FG02-07ER46471.
\par
Eric Pop acknowledges support from the Marco MSD center and ONR-YIP grant no.\ N00014-10-1-0853.
\par
Part of this work has been carried on under the contract no.\ 34524/2007 of the Intel Corporation, whose support is gratefully acknowledged.

\appendix
\section*{Appendix -- Derivation of the optical properties from the band
diagram}\label{optical_equations} 
\setcounter{section}{1}
In the framework of band theory without electron-hole interaction, the dielectric tensor $\varepsilon_{\alpha \beta}$ is defined as
\begin{eqnarray}\label{dielectrictensor}
\fl
\varepsilon_{\alpha \beta}(\omega)=
1 + \frac{e^2}{\varepsilon_0 \Omega m^2}
\sum_{n,n'}\sum_{\bf k}
\frac{\Mbf_{\alpha \beta}^{nn'} [f(E_{{\bf k}n}) - f(E_{{\bf k}n'})\big]} {(E_{{\bf k}n'}-E_{{\bf k}n})^2}+ \nonumber \\
+ \left[ 
\frac{1}{(\omega_{{\bf k}n'}-\omega_{{\bf k}n})+\omega + \rmi\Gamma\omega  } +
\frac{1}{(\omega_{{\bf k}n'}-\omega_{{\bf k}n})-\omega - \rmi\Gamma\omega  } 
\right]
\end{eqnarray}
with $\Gamma \!\! \rightarrow \!\! 0^+$. In (\ref{dielectrictensor}) $e$, $m$ and $\Omega$ are the electron charge and mass, and the volume of the lattice cell, respectively; $E_{{\bf k}n}$ are the eigenvalues of the Hamiltonian and $f(E_{{\bf k}n})$ is the Fermi distribution function accounting for the band occupation. Letting 
$
\omega_p=\sqrt{
({e^2 N})/{(\varepsilon_0 m)}}$
be the plasma frequency with $N$ standing for the number of electrons per unit volume, and 
$
\Delta=\big[(\omega_{{\bf k}n'}-\omega_{{\bf k}n})^2-\omega^2\big]^2+\Gamma^2\omega^2
$, the imaginary part ${\epsilon_i}_{\alpha \beta}(\omega)$  of  the  dielectric tensor
$\epsilon_{\alpha \beta}   (\omega)$  is   given  by   the   following  Drude-Lorentz
expression:
\begin{eqnarray}
\fl
{\epsilon_i}_{\alpha \beta} (\omega) = 
\frac{\omega_p^2}{Nm\Omega} 
\left[
\sum_{n, \bf{k}}
\frac{ \ddri f(E_{{\bf k}n}) }{\ddri E_{{\bf k}n} } \, 
\frac{ \eta \, \omega \Mbf_{\alpha \beta}^{nn'}} {\omega^4+\eta^2 \omega^2} + \right. \nonumber \\
 \left. +2 \sum_{n,n'}\sum_{\bf k}
\frac{f(E_{{\bf k}n})}{E_{{\bf k}n'}-E_{{\bf k}n}}
\frac{\Gamma \, \omega \Mbf_{\alpha \beta}^{nn'}}{\Delta}
\right] \,,
\label{epsiloni}
\end{eqnarray}
where the original sum over $n$ and $n'$ of (\ref{dielectrictensor}) has been split into two terms, the former accounting for valence-to-valence (or conduction-to-conduction) intraband transitions ($n'=n$), the latter standing for transitions from states belonging to the valence band (index $n$) to states belonging to the conduction band (index $n'$). In the summands, the squared matrix elements $\Mbf_{\alpha \beta}^{nn'}$ are weighted by a smearing coefficient ($\eta$ or $\Gamma$), and by a factor depending on the Fermi distribution function for interband transitions, or on its derivative for the intraband contribution. Considering that the derivative is substantially zero except in the region close to the Fermi level, the dielectric tensor is dominated by interband transitions, as expected. Nevertheless, a few states near the top of the valence band can be empty due to thermal excitations and, conversely, a small amount of states in the conduction band are occupied. As a consequence, a number of intraband transitions occur, that are described by the first summand of equations (\ref{epsiloni}) and (\ref{epsilonr}). Accounting for such transitions is useful to better reproduce the experimental behaviour.
\par
In order to keep the Drude-Lorentz approximation valid, the two smearing coefficients $\eta$ and $\Gamma$ must be small, even though not vanishing. For the case described in the text they were treated as fitting parameters and both set to 1.0 for the hexagonal phase and to 0.8 and 0.3, respectively, for the fcc phase.
 \par
The real ${\epsilon_r}_{\alpha \beta}(\omega)$ part of the dielectric tensor is then calculated applying the Kramers-Kronig relationship to (\ref{epsiloni}):
\begin{eqnarray}
\fl
{\epsilon_r}_{\alpha \beta}(\omega)=
1- \frac{\omega_p^2}{Nm\Omega} 
\left[
\sum_{n, {\bf k}}
\frac{ \ddri f(E_{{\bf k}n}) }{\ddri E_{{\bf k}n} } \, 
\frac{\omega^2 \Mbf_{\alpha \beta}^{nn'}} {\omega^4+\eta^2 \omega^2}+ \right. \nonumber\\
\left. -2 \sum_{n,n'}\sum_{\bf k}
\frac{f(E_{{\bf k}n})}{E_{{\bf k}n'}-E_{{\bf k}n}}
\frac{(\Delta-\Gamma^2 \, \omega^2) \Mbf_{\alpha \beta}^{nn'} } {\Delta} 
\right] \,.
\label{epsilonr}
\end{eqnarray}
The squared matrix elements $\Mbf_{\alpha \beta}^{nn'}$ reveals the tensorial nature of $\varepsilon_{\alpha \beta}(\omega)$ and are defined as follows: 
\begin{equation}
\Mbf_{\alpha \beta}^{nn'}=\langle u_{{\bf k}n'} \vert {\bf p}_{\alpha} \vert u_{{\bf k}n} \rangle
\langle u_{{\bf k}n} \vert {\bf p}^{\dagger}_{\beta} \vert u_{{\bf k}n'} \rangle
\label{nos}
\end{equation}
where $\vert u_{{\bf k}n} \rangle$ is a factor of the single particle Bloch function obtained by the Kohn-Sham DFT calculation, and ${\bf p}_{\alpha}$ is the momentum operator along the $\alpha$ direction.
\par
In a principal system, the off-diagonal elements of the dielectric tensor are zero and, for perfectly isotropic materials, the diagonal elements are equal. For the two systems considered here, only two eigenvalues out of three are equal. In order to compare results with experimental data where isotropy is assumed, the eigenvalues of the dielectric tensor have been averaged to obtain a unique function.
\par
The refractive index $n(\omega)$, the extinction coefficient $k(\omega)$ and the absorption coefficient $\alpha(\omega)$ are calculated by means of the Maxwell model through the following relationships:
\begin{eqnarray}
\label{n}
n (\omega)&=&\sqrt{\frac{\sqrt{\epsilon_r(\omega)^2 + \epsilon_i(\omega)^2}+\epsilon_r(\omega)}{2}},\\
\label{k}
k (\omega)&=&\sqrt{\frac{\sqrt{\epsilon_r(\omega)^2 + \epsilon_i(\omega)^2}-\epsilon_r(\omega)}{2}},\\
\label{a}
\alpha (\omega)&=&\frac{\omega}{cn(\omega)}\epsilon_i(\omega)=\frac{2\omega}{c}k(\omega). 
\end{eqnarray}
where the symbols $\epsilon_r$ and $\epsilon_i$ without superscripts represent an average function determined as described above.
\end{document}